\documentclass[acmtog,authorversion,screen]{acmart}
\AtBeginDocument{%
  \providecommand\BibTeX{{%
    \normalfont B\kern-0.5em{\scshape i\kern-0.25em b}\kern-0.8em\TeX}}}
\setcopyright{acmcopyright}
\copyrightyear{2024}
\acmYear{2024}
\acmDOI{XXXXXXX.XXXXXXX}
\acmSubmissionID{181}
\usepackage{stfloats}
\usepackage{float} 
\usepackage{subfigure}  
\usepackage{algorithm}
\usepackage{algorithmic}
\usepackage{bm}
\usepackage{multirow}
\usepackage{xcolor}
\usepackage{amsmath}
\usepackage{hyperref} 
\hypersetup{colorlinks = true,
	    linkcolor = black,
	    urlcolor = blue,
            citecolor = blue}
\newcommand{\note}[1]{#1}

\newcommand{\mbx}{\boldsymbol{x}}
\newcommand{\mbp}{\boldsymbol{p}}
\newcommand{\mbP}{\boldsymbol{P}}
\newcommand{\mbe}{\boldsymbol{e}}
\newcommand{\mbn}{\boldsymbol{n}}

\newcommand{\mbh}{\boldsymbol{h}}
\newcommand{\mbN}{\boldsymbol{N}}

\newcommand{\mbt}{\boldsymbol{t}}

\newcommand{\mbb}{\boldsymbol{b}}

\newcommand{\mbT}{\mathcal{T}}
\newcommand{\mbd}{\boldsymbol{d}}
\newcommand{\mbu}{\boldsymbol{u}}

\author{Zhimin Fan}
\email{zhiminfan2002@gmail.com}
\affiliation{%
  \institution{State Key Lab for Novel Software Technology, Nanjing University}
  \city{Nanjing}
  \country{China}
}
\author{Jie Guo}
\email{guojie@nju.edu.cn}
\authornote{Corresponding author.}
\affiliation{%
  \institution{State Key Lab for Novel Software Technology, Nanjing University}
  \city{Nanjing}
  \country{China}
}
\author{Yiming Wang}
\email{02yimingwang@gmail.com}
\author{Tianyu Xiao}
\email{tianyuxiaoty@outlook.com}
\affiliation{%
  \institution{State Key Lab for Novel Software Technology, Nanjing University}
  \city{Nanjing}
  \country{China}
}
\author{Hao Zhang}
\email{213210060@seu.edu.cn}
\affiliation{%
  \institution{Southeast University}
  \city{Nanjing}
  \country{China}
}
\author{Chenxi Zhou}
\email{502023320017@smail.nju.edu.cn}
\author{Zhenyu Chen}
\email{chenzy@smail.nju.edu.cn}
\affiliation{%
  \institution{State Key Lab for Novel Software Technology, Nanjing University}
  \city{Nanjing}
  \country{China}
}
\author{Pengpei Hong}
\email{hpommpy@gmail.com}
\affiliation{%
  \institution{University of Utah}
  \city{Salt Lake City} 
  \country{United States of America}
}
\author{Yanwen Guo}
\email{ywguo@nju.edu.cn}
\affiliation{%
  \institution{State Key Lab for Novel Software Technology, Nanjing University}
  \city{Nanjing}
  \country{China}
}
\author{Ling-Qi Yan}
\email{lingqi@cs.ucsb.edu}
\affiliation{%
  \institution{University of California, Santa Barbara}
  \city{Santa Barbara}
  \country{United States of America}
}

\begin{document}

\setcopyright{acmlicensed}
\acmJournal{TOG}
\acmYear{2024} \acmVolume{43} \acmNumber{4} \acmArticle{1} \acmMonth{8} \acmPrice{15.00}\acmDOI{10.1145/3618360}
\title{Specular Polynomials}

\begin{teaserfigure}
\includegraphics[width=0.994\textwidth]{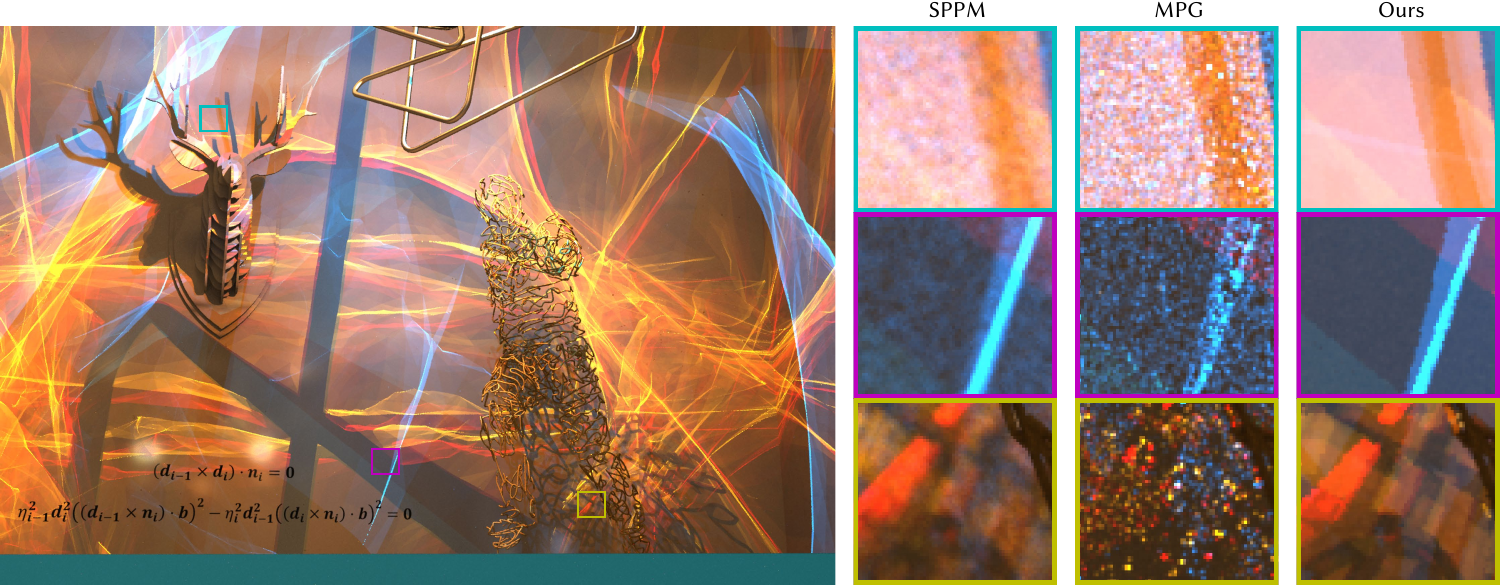}  
  \caption{We render a shop window scene featuring challenging caustics and complex visibility, using our pipeline based on specular polynomials. The caustics stem from colored point light sources placed inside a dielectric object, and the whole scene is viewed through a transparent window. Such a configuration makes most existing rendering algorithms fail, while our method succeeds in reproducing the stunning light transport effect.
 The insets show equal\note{-}time (10 min) comparisons against Stochastic Progressive Photon Mapping (SPPM) \cite{Hachisuka09SPPM} and Manifold Path Guiding (MPG) \cite{Fan23MPG}.}
  \Description{...}
  \label{fig:teaser}
\end{teaserfigure}
\begin{abstract}

Finding valid light paths that involve specular vertices in Monte Carlo rendering requires solving many non-linear, transcendental equations in high-dimensional space. Existing approaches heavily rely on Newton iterations in path space, which are limited to obtaining at most a single solution each time and easily diverge when initialized with improper seeds.

We propose \textit{specular polynomials}, a Newton iteration-free methodology for finding a complete set of admissible specular paths connecting two arbitrary endpoints in a scene. 
The core is a reformulation of specular constraints into polynomial systems, which makes it possible to reduce the task to a univariate root-finding problem.
We first derive bivariate systems utilizing rational coordinate mapping between the coordinates of consecutive vertices. Subsequently, we adopt the hidden variable resultant method for variable elimination, converting the problem into finding zeros of the determinant of univariate matrix polynomials.
This can be effectively solved through Laplacian expansion for one bounce and a bisection solver for more bounces.

Our solution is generic, completely deterministic, \note{accurate for the case of one bounce}, and GPU-friendly. We develop efficient CPU and GPU implementations and apply them to challenging glints and caustic rendering. Experiments on various scenarios demonstrate the superiority of specular polynomial-based solutions compared to Newton iteration-based counterparts. \note{Our implementation is available at \href{https://github.com/mollnn/spoly}{https://github.com/mollnn/spoly}.}

\end{abstract}

\ccsdesc[500]{Computing methodologies~Ray tracing}

\keywords{Specular chain, Polynomial, Caustics, Glints}

\maketitle

\section{Introduction}

Although Monte Carlo (MC) rendering algorithms have made enormous strides in reducing noise \cite{Christensen16, Fascione18, Keller15}, they are still inadequate \note{for handling} certain kinds of light paths. Paths containing specular chains (i.e., multiple consecutive specular scattering events) are extremely difficult to sample for existing MC rendering algorithms. The bottleneck is that a light path satisfying all physical constraints at specular reflective/refractive vertices has an infinitely small probability for sampling. To alleviate this issue, prior methods resort to some kind of multivariate Newton solver: seed paths are heuristically generated or randomly sampled first and then undergo Newton iterations in path space (e.g., manifold walk) to reach admissible paths \cite{Jakob12, Zeltner20}.

Unfortunately, Newton solvers do not always converge and are highly sensitive to the selection of initial seed paths. Improper seed paths can lead to divergence and hence introduce substantial bias or variance to the final rendering \cite{Hanika15, Zeltner20, Wang20}. Several recent works design dedicated strategies to carefully and intelligently select seed paths leveraging historical samples \cite{Fan23MPG, Yu2023NeuralPS, xu2023efficient}. Even so, missing high-frequency optical details (e.g., glints and caustics) and producing severe outliers happen frequently in complex scenes. Moreover, Newton solvers that walk on the whole specular manifolds \cite{Jakob12} will also fail when facing scenes containing small specular geometries and complex visibility \cite{Otsu2018GeometryawareML}, since they rely heavily on local continuity.

\begin{figure}
    \centering
    \includegraphics[width=\linewidth]{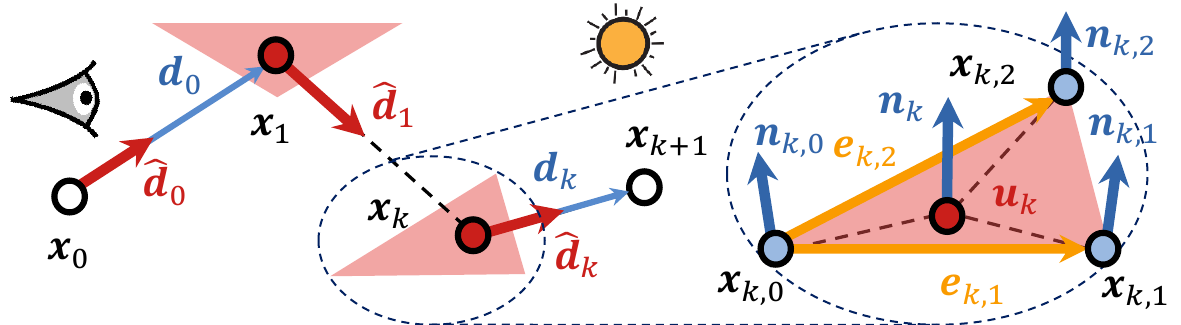}
    \caption{Illustration of our problem setup and important symbols.}
    \label{fig_setup}
\end{figure}

In this paper, we endeavor to remove the Newton solver from specular chain sampling, thereby fundamentally avoiding the aforementioned issues. Our key insight is that Newton\note{'s} iteration-based methods rely solely on information about a single point at a time. While this aligns seamlessly with manifold-based techniques \cite{Jakob12, Hanika15, Zeltner20}, there are also quite a few cases in which we possess comprehensive information about an entire region. Specifically, when considering paths passing through a tuple of triangles in Fig. \ref{fig_setup}, we can express the specular constraints in closed-form equations using vertex positions and normals \cite{Walter09, Wang20}. Can we therefore obtain all of the solutions directly from these closed-form equations?
 
To this end, we derive \textbf{specular polynomials}, a polynomial formulation of specular constraints, and solve these polynomials directly, bypassing the use of multivariate Newton's method. Here, a significant challenge arises due to the high dimensionality involved. A specular chain with $k$ vertices requires solving the problem in a space of $2k$ dimensions. To reduce the number of variables, we recursively apply rational mapping to vertex coordinates along a specular chain, resulting in \textbf{bivariate specular polynomials}. Leveraging the hidden variable resultant method tailored for polynomial equations \cite{nakatsukasaComputingCommonZeros2015}, we further convert them into \textbf{univariate specular polynomials}. This allows us to adopt a wide variety of methods in mathematics, such as root isolation \cite{Collins76}, bisection, and eigenvalue decomposition \cite{Golub2012MatrixComputation}, to effectively solve the simplified problem with only one variable.

When applied to the rendering of glints and caustics, our \note{solver can be run on the triangle tuples selected using existing techniques \cite{Wang20} that prune non-contributing regions of the path space}. This pipeline can simulate these challenging effects, consuming even less time than previous methods for one bounce. As a direct and deterministic method, it naturally handles discontinuities raised by small specular geometries and complex visibility, producing fewer artifacts such as strong outliers and energy loss.

In summary, our main contributions include:

\begin{itemize}
  \item \textbf{A polynomial formulation of specular constraints}, derived by combining vertex constraint polynomials and rational coordinate mappings between barycentric coordinates.
  \item \textbf{A specular path solver} using hidden variable resultant method combined with direct or eigenvalue solvers, which is deterministic and free from multivariate Newton iterations. 
  \item \textbf{Applications to glints and caustics rendering}, which achieves fast and almost noise-free rendering of specular light transport effects.
\end{itemize}

\section{Related Works}

\paragraph{Specular light transport}

Since caustics and glints can not be robustly handled by (bidirectional) path tracing \cite{Kajiya86, Veach95BDPT} even when properly guided \note{\cite{Muller17, Vorba19, Ruppert2020RobustFO, Rath23, Reibold2018SelectiveGS}} or using Metropolis sampling \cite{Veach97MLT}, numerous specialized rendering methods have emerged.
One line of approaches involves searching and root-finding techniques to identify all specular chains connecting two endpoints. \citet{Mitchell92} utilize Fermat's principle with interval Newton's method to identify specular reflection paths including non-specular vertices. 
\note{\citet{Chen00} examine the reflection geometry by employing Fermat’s principle to establish a path Jacobian and path Hessian, which addresses perturbations in the endpoint of a path and enhances the rendering of reflections on curved surfaces.}
\citet{Walter09} introduce a pruning strategy and employ Newton's method to discover refractive specular paths. \citet{Loubet20} analytically estimate the contribution of each specular triangle under far-field assumptions, enabling importance sampling of specular vertices, though requiring recursive subdivision to reduce bias for pure specular cases. However, these methods are tailored for a single specular bounce. \citet{Wang20} extend \citet{Walter09}'s method to multi-bounce specular chains. To mitigate performance degradations, they resort to using the regular Newton's method instead of the interval variant, sacrificing some solutions for computational efficiency and practicality. 

Alternatively, other methods alleviate the computational burden through stochastic sampling. Manifold Exploration Metropolis Light Transport (MEMLT) \cite{Jakob12} facilitates random walks on a specular manifold using Newton's method, which then extends to the half-vector space \cite{Kaplanyan2014} and integrates into regular Monte Carlo sampling as Manifold Next-Event Estimation (MNEE) \cite{Hanika15}. \note{MNEE augments both unidirectional and bidirectional techniques \cite{Speierer2018CausticCS}.} However, with fixed initialization, MNEE can only identify at most one specular chain connecting a given pair of endpoints, resulting in energy loss. Specular Manifold Sampling (SMS) \cite{Zeltner20} tackles this challenge by employing random initialization, coupled with an unbiased reciprocal probability estimator. \note{The convergence rate of SMS is subsequently improved by introducing a large jump \cite{jhang2022specular}.}
\note{Additionally, for the volumetric counterparts, \citet{pediredlaPathTracingEstimators2020} and \citet{fraboniCanYouSee2023} apply gradient-based optimizations to the problem of refractive radiative transfer.}
Nevertheless, the variance remains unbounded, as the convergence basin for each solution can become uncontrollably small, even when properly guided by reconstructed importance seed distributions \cite{Fan23MPG}.

In stark contrast to the gradient-based iteration methods that operate in path space, we introduce the first generic framework for computing all solutions without reliance on multivariate Newton iterations. The proposed framework combines analytical derivation with numerical computation, robustly handling the intricacies of specular constraints.

\paragraph{Specialized methods for rendering glints and caustics}

Industrial caustics rendering has conventionally employed photon-based approaches \cite{Jensen95PM, Hachisuka08PPM, Hachisuka09SPPM} or regularization \cite{Kaplanyan2013PathSR, Jendersie2019MicrofacetMR, Weier2021OPSR}. However, the inherent bias caused by spatial relaxation in these methods frequently results in issues such as light leaking and unexpected blurring, while our method solves the original, unrelaxed problem.

Specialized glint-rendering methods are mostly limited to a single reflection event and are designed exclusively for surfaces with normal maps \cite{Yan14, Zhu22GlintySurvey}. Alternatively, some works concentrate on rendering procedurally generated glints \cite{Jakob2014Discrete}. In contrast, our approach handles actual geometric primitives, accommodating multiple bounce reflections and refractions, which successfully addresses both glints and caustic rendering tasks.

\paragraph{Polynomial root-finding techniques}

Many mathematical tools for elimination and root-finding are tailored for polynomials. For univariate polynomials, real root isolation \cite{Collins76} by differentiation can be used to develop robust solvers \cite{yukselHighPerformancePolynomialRoot2022}. Another approach involves solving the eigenvalues of \note{the companion matrices} through QR decomposition \cite{Golub2012MatrixComputation}.

Elimination methods can be employed to handle polynomial systems involving two or more variables \cite{buchbergerGrobnerBasesIntroduction1992, hiltonMarchingTrianglesRange1996}. 
The resultant method is efficient for discovering all solutions of multivariate polynomials systems \cite{nakatsukasaComputingCommonZeros2015}, which is widely embraced in many fields \cite{ sadeghimaneshResultantToolsParametric2022, kapurAlgebraicGeometricReasoning1994, kajiyaRayTracingParametric1982}.

Substantial research has focused on the resultant. Various resultant matrices, such as Sylvester \cite{Sylvester1853} and B\'{e}zout \cite{Bezout1779} for bivariate polynomial systems and Dixon \cite{Dixon1908} for multivariate (three or more) polynomial systems, have been explored. Fast algorithms with low time complexity for computing the resultant matrix have been developed \cite{chionhFastComputationBezout2002, emirisImprovedAlgorithmsComputing2005, grenetComplexityMultivariateResultant2013, qinComplexityConstructingDixon2017}. 


In this work, we formulate the constraints as polynomials for elimination. We employ the B\'{e}zout resultant for bivariate polynomial systems predominantly due to its numerical stability and low computational complexity \cite{chionhFastComputationBezout2002}. 
\note{
Although existing works also mention some polynomial forms of specular constraints \cite{Glaeser00}, they are in a very different context.
}

\section{Polynomial Forms of Specular Constraints}

In this section, we convert the specular constraints into polynomial systems, starting with the definition of our problem setup.

\subsection{Problem definition} 

\begin{table}[tbp]
\centering
\caption{List of important symbols. By default, a vector with a hat means it is normalized.}
\label{tab:parameters}
\begin{tabular}{ll}
\toprule
\textbf{Symbol} & \textbf{Description} \\ 
\midrule
$\mbx_0, \mbx_{k+1}$ & Position of non-specular separators \\
$\mbx_i$ & Position of specular vertices \\ 
$\mbP_{i}$ & Position matrix $(\mbp_{i,0}, \mbp_{i,1}, \mbp_{i,2})$ \\ 
$\mbe_{i,1}, \mbe_{i,2}$ & Vector of triangle edges \\ 
$\mbn_i$ & Un-normalized linearly interpolated normal of $\mbx_i$ \\ 
$\hat{\mbn}_i$ & Normal vector of $\mbx_i$ \\ 
$\mbN_{i}$ & Normal matrix $(\mbn_{i,0}, \mbn_{i,1}, \mbn_{i,2})$ \\ 
$\mbh_i$ & Generalized half-vector of $\mbx_i$ \\ 
${\mbt}_{i,1}, {\mbt}_{i,2}$ & Tangent vectors of $\mbx_i$, computing from ${\mbn}_i$ and $\mbe_{i,1/2}$\\ 
$\mbd_i$ & Position difference of vertices $\mbx_{i+1}$ and $\mbx_{i}$ \\ 
$\hat{\mbd}_i$ & Direction from $\mbx_{i}$ to $\mbx_{i+1}$ \\ 
$\mbu_i$ & Barycentric coordinate of  $\mbx_i$\\ 
\bottomrule
\end{tabular}
\label{tab_sym}
\end{table}

Formally, we denote two fixed separators as $\mbx_0$ and $\mbx_{k+1}$, where $\mbx_0$ \note{could be a vertex on the camera or a non-specular shading point, and $\mbx_{k+1}$ could be a vertex on the light source or another non-specular shading point in bidirectional techniques \cite{Jakob12, Fan23MPG, Speierer2018CausticCS}}. 
A specular chain connecting $\mbx_0$ and $\mbx_{k+1}$ is represented by $\overline{\mbx}$, which is comprised of specular vertices $\mbx_1, \mbx_2, \ldots, \mbx_k$. Important symbols used in this paper are summarized in Table \ref{tab_sym} and illustrated in Fig. \ref{fig_setup}.

\note{Existing pruning techniques \cite{Walter09, Wang20} can be used to select the triangle tuples that may contain a specular chain connecting the two separators. Thus, our problem begins with a given tuple of} $k$ triangles $\mbT_1, \ldots, \mbT_k$, each consisting of three vertices with positions\footnote{Here, we simplify the representation by using matrices (position matrix $\mbP_i$ and normal matrix $\mbN_i$) to express the positions of the three vertices of a triangle. Thus, interpolations in Eq. (\ref{eq_xi}) and Eq. (\ref{eq_ni}) can be written as matrix-vector multiplications.} $\mbP_i=\left(\mbp_{i,0}, \mbp_{i,1}, \mbp_{i,2}\right)$ and normals $\mbN_i=\left(\mbn_{i,0}, \mbn_{i,1}, \mbn_{i,2}\right)$, our objective is to determine the set of all possible tuples of specular vertices $(\mbx_1, \ldots, \mbx_k)$ that forms admissible specular chains, where each vertex $\mbx_i$ lies on a triangle $\mbT_i$ with interpolated normal vectors $\hat{\mbn}_i$. To make the problem tractable, we resort to the barycentric coordinates $\mbu_i=\left(1-u_i-v_i, u_i,v_i\right)^\top$ to represent the positions of these vertices:
\begin{equation}
\mbx_i = (1-u_i-v_i) \mbp_{i,0} + u_i \mbp_{i,1} + v_i \mbp_{i,2} = \mbP_{i} \mbu_i.
\label{eq_xi}
\end{equation}
Similarly, the normal vector $\hat{\mbn}_i$ of specular vertices is also determined by the barycentric interpolation from the vertex normals:
\begin{equation}
\hat{\mbn}_i = \frac{{\mbn_i}}{   \Vert {\mbn_i} \Vert}, \quad 
{\mbn}_i = (1-u_i-v_i) \mbn_{i,0} + u_i \mbn_{i,1} + v_i \mbn_{i,2} = \mbN_{i} \mbu_i.
\label{eq_ni}
\end{equation} 
Here, $\mbn_i$ is un-normalized and is linear to $\mbu_i$. Normalizing $\mbn_i$ results in $\hat{\mbn}_i$ which is no longer linear to $\mbu_i$.
For simplicity, we introduce dummy barycentric coordinates $\mbu_0$ and $\mbu_{k+1}$ for separators. Eq. (\ref{eq_xi}) still holds for $\mbp_{0,0}=\mbp_{0,1}=\mbp_{0,2}=\mbx_{0}$ and $\mbp_{k+1,0}=\mbp_{k+1,1}=\mbp_{k+1,2}=\mbx_{k+1}$, and $\mbu_0$ and $\mbu_{k+1}$ will not appear in the final equations.

Ideal specular reflection or refraction occurs at each vertex, imposing constraints on the positions of consecutive vertices and their normal vectors. Specifically, for each specular vertex $\mbx_i$, the constraint can be characterized by \cite{Kaplanyan2014, Hanika15, Wang20, Zeltner20}
\begin{equation}
\mbh_i \times \mbn_i=\boldsymbol{0},1\le i\le k.
\label{eq_ch}
\end{equation}
Here, $\mbh_i = \eta_i \hat{\mbd}_{i} - \eta_{i-1}\hat{\mbd}_{i-1}$ is the generalized half vector \cite{Walter2007MicrofacetMF},
where 
\note{\begin{equation}
\hat{\mbd}_i=\frac{\mbd_i} {\Vert \mbd_i \Vert}, \quad \mbd_i= \mbx_{i+1} - \mbx_i,
\end{equation}}
and $\eta_i$ represents the index of refraction of the outgoing side of $\mbx_i$. Both $\mbh_i$ and $\mbn_i$ vary with respect to $u_i$ and $v_i$.

Our problem is to solve the above constraint equations for variables $\mbu_i = (u_i, v_i)$, which forms a system of $2k$ independent equations on $2k$ variables $u_i$ and $v_i$ ($i=1,\ldots, k$). \note{This has been proven to have only a finite number of solutions \cite{Wang20}.} Unfortunately, these equations are not polynomials due to the existence of square roots and fractions. Thus, the rest of this section is devoted to transforming them into specular polynomials. Furthermore, we will attempt to maintain the degree and number of variables as low as possible, as this is critical to the performance of the solving process.

\subsection{Vertex constraints}

To convert the aforementioned constraint equations into polynomials, we first consider expressing the specular constraint on vertex $\mbx_i$ as a polynomial in variables $\mbu_{i-1}$, $\mbu_i$, and $\mbu_{i+1}$. Because these constraints solely deal with the reflection/refraction behavior on a single vertex, they are referred to as \textbf{vertex constraints}.
To clarify the derivation, we treat vertex constraints as two separate conditions: the coplanarity constraint and the angularity constraint.

\paragraph{Coplanarity constraint}

The reflection/Snell's law states that the normal vector $\hat{\mbn}_i$, the outgoing direction $\hat{\mbd}_i$, and the incident direction $\hat{\mbd}_{i-1}$ all reside in the same plane, i.e.,
\begin{equation}
\boxed{
(\mbd_{i-1}\times \mbd_{i}) \cdot \mbn_i = 0.
}
\label{eq_coplane1}
\end{equation}
Note that we have excluded the normalization factors, so this is already a polynomial equation in $\mbu_{i-1}$, $\mbu_i$, and $\mbu_{i+1}$.
We can simplify it further using $\mbd_{i-1}+\mbd_{i}=\mbx_{i+1}-\mbx_{i-1}$ and $\mbd_{i-1} \times \mbd_{i-1} = \boldsymbol{0}$:
\begin{equation}
\boxed{
(\mbd_{i-1} \times
(\mbx_{i+1} - \mbx_{i-1})
) \cdot 
{\mbn}_i 
= 0.
}
\label{eq_coplane}
\end{equation}

\paragraph{Angularity constraint} The incident and reflected/refracted angle values must satisfy the reflection/Snell's law:
\begin{equation}
\eta_{i-1}  \Vert \hat{\mbd}_{i-1} \times \mbn_i  \Vert = \eta_{i} \Vert \hat{\mbd}_{i} \times \mbn_i \Vert.
\label{eq_angrel}
\end{equation}
\note{Since the coplanarity condition implies that $\hat{\mbd}_{i-1}$, $\hat{\mbd}_{i}$, and $\mbn_i$ lies on the same plane, $\hat{\mbd}_{i-1} \times \mbn_i$ and $\hat{\mbd}_{i} \times \mbn_i$ are parallel to each other, i.e.,
\begin{equation}
\frac{
 \hat{\mbd}_{i-1} \times \mbn_i  
}{\Vert \hat{\mbd}_{i-1} \times \mbn_i  \Vert}
=
\frac{
 \hat{\mbd}_{i} \times \mbn_i  
}{\Vert \hat{\mbd}_{i} \times \mbn_i  \Vert}.
\end{equation}
Therefore, Eq. (\ref{eq_angrel}) becomes}
\begin{equation}
\eta_{i-1}  \hat{\mbd}_{i-1} \times \mbn_i   = \eta_{i}  \hat{\mbd}_{i} \times \mbn_i
\label{eq_angrel2}
\end{equation} 
with $\eta_{i} = \eta_{i-1}$ handling the reflection case.

Unfortunately, the square roots in $\hat{\mbd}_{i-1}$ and $\hat{\mbd}_{i}$ indicate that Eq. (\ref{eq_angrel2}) is not a polynomial. We propose two polynomialization techniques to resolve this issue.

\subsection{Polynomialization}

Making \note{the aforementioned angularity constraint} a polynomial requires removing the square roots in the denominators. To achieve this goal, we propose two \note{ways} by computing squares or constructing common factors, respectively: \note{The \textbf{square form} handles both reflection and refraction, but the \textbf{product form}, which has a lower degree than the square form, only works for reflection. Therefore,} these two forms are used jointly in our final pipeline. 

\paragraph{Square form} 

A straightforward way to remove the square roots is by performing squaring. 
To convert it into a scalar equation, we project\footnote{Projection means computing the dot product of Eq. (\ref{eq_angrel2}) with the basis $\mbb$. It should also be noted that Eq. (\ref{eq_cpf_result}) is merely a necessary condition of Eq. (\ref{eq_angrel2}). Yet, even if some superfluous solutions are discovered, they will be immediately filtered out after checking the specular constraints in path space. The same goes for Eq. (\ref{eq_pf_result}).} Eq. (\ref{eq_angrel2}) onto an arbitrary \note{vector} $\mbb$. Then, we eliminate the square roots by squaring both sides and remove the denominators by multiplying the common denominator $\mbd_i^2 \mbd_{i-1}^2$, resulting in\footnote{\note{Here, we use the notation $\mbx^2$ to represent the dot product $\mbx \cdot \mbx$ of a vector $\mbx$.}}:
\begin{equation}
\boxed{
\eta_{i-1}^2 {{{\mbd}_{i}^2}} \left({({\mbd}_{i-1}\times {\mbn}_i) \cdot \mbb}\right)^2
=
\eta_i^2 {{{\mbd}_{i-1}^2}} \left({({\mbd}_{i}\times {\mbn}_i) \cdot \mbb}\right)^2.
}
\label{eq_cpf_result}
\end{equation} 
This equation represents a polynomial in terms of $\mbu_{i-1}$, $\mbu_i$, and $\mbu_{i+1}$.
In practice, we choose $\mbb = (1, 0, 0)^\top$ by default and switch to $(0, 0, 1)^\top$ if the equation degenerates. \note{Note that the square operations may introduce additional solutions with positive and negative signs, so we have to check the original constraints, Eq. (\ref{eq_ch}), to remove such superfluous solutions.}


\paragraph{Product form} 

\begin{figure}
    \centering
    \includegraphics[width=\linewidth]{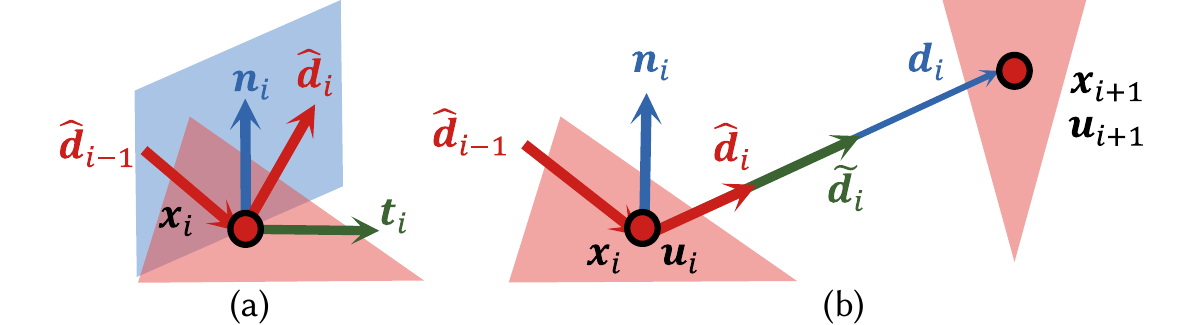}

    \caption{(a) Product form of vertex constraints requires decomposing the ray in the local tangent frame of $\mbx_i$. (b) Recursive rational mappings represent $\mbu_{i+1}$ using a rational expression of $\mbu_i$. }
    \label{fig_proj}
\end{figure}

An alternative method is to remove the normalization operation directly. This requires constructing two equations both with factors $\Vert \mbd_{i-1} \Vert$ and $\Vert \mbd_{i} \Vert$. As illustrated in Fig. \ref{fig_proj}(a), for a reflective vertex $\mbx_i$, we can exploit the symmetry property that the angle between $\hat{\mbd}_{i-1}$ and $\mbn_i$ equals the angle between $-\hat{\mbd}_{i}$ and $\mbn_i$:
\begin{equation}
\hat{\mbd}_{i-1} \cdot \mbn_i = -\hat{\mbd}_{i} \cdot \mbn_i.
\end{equation}
Supposing that an arbitrary direction $\mbt_i$ is perpendicular to $\mbn_i$, the angle between $\hat{\mbd}_{i-1}$ and $\mbt_i$ also equals the angle between $\hat{\mbd}_{i}$ and $\mbt_i$:
\begin{equation}
\hat{\mbd}_{i} \cdot \mbt_i = \hat{\mbd}_{i-1} \cdot \mbt_i.
\end{equation}
These equations can be viewed as decomposing the incident and outgoing directions in the local tangent frame of $\mbx_i$. 
Afterward, to eliminate the square root in the normalization factors of $\hat{\mbd}_{i-1}$ and $\hat{\mbd}_i$ in the above equations, we multiply them together.
This results in a polynomial equation in terms of $\mbu_{i-1}$, $\mbu_i$, and $\mbu_{i+1}$:
\begin{equation}
\boxed{
(\mbd_{i-1} \cdot {\mbn}_i) (\mbd_{i} \cdot {\mbt}_i) + (\mbd_{i-1} \cdot {\mbt}_i) (\mbd_{i} \cdot {\mbn}_i) = 0.
}
\label{eq_pf_result}
\end{equation}
In practice, we choose ${\mbt}_i = \mbn_i \times \mbe_{i,1}$ by default, where $\mbe_{i,1} = \mbp_{i,1} - \mbp_{i,0}$. We switch to ${\mbt}_i = \mbn_i \times \mbe_{i,2}$ if the equation degenerates.

\paragraph{Multivariate specular polynomials}


Using the above two polynomialization algorithms, we can directly construct a multivariate polynomial system with $2k$ variables $\mbu_1, \ldots, \mbu_k$. We refer to this system as \textbf{multivariate specular polynomials}. Table \ref{tab_equ} summarizes these polynomials. The maximum degree is $6$ for refraction and $4$ for reflection \note{when using interpolated shading normals. Our proposed formulation still applies to specular triangles with face normals, where $\mbn_i = \mbn_{i,0}$ is a constant. In this case, the maximum degree is $4$ for refraction and $2$ for reflection.}

Due to numerical instability and computational burden \cite{noferiniNumericalInstabilityResultant2016, qinComplexityConstructingDixon2017}, directly solving these multivariate formulations with many variables is not recommended. 
In mathematics, the most reliable solvers for polynomial systems are designed for two variables \cite{nakatsukasaComputingCommonZeros2015}. Inspired by this, we further transform all multivariate formulations into bivariate ones, making the method much more practicable. Fortunately, this conversion can be accomplished through a change of variables leveraging the light transport behavior of specular paths.


 




\begin{table*}[tbp]
\centering
\caption{Summary of vertex constraint polynomials. \note{We show the degree when using interpolated shading normals (I) and face normals (F).}}
\begin{tabular}{l|lr|lr}
\toprule
\textbf{Constraints} & \textbf{Formulation I} & Degree & \textbf{Formulation II} & Degree \\ 
& Equation & \note{I/F} & Equation & \note{I/F} \\
\midrule
Coplanarity & Consecutive difference form & 3\note{/2} & Endpoint difference form & 2\note{/1} \\
& $(\mbd_{i-1}\times \mbd_{i}) \cdot \mbn_i = 0$ & & ($\mbd_{i-1} \times
(\mbx_{i+1} - \mbx_{i-1})
) \cdot 
{\mbn}_i 
= 0$ & \\
\midrule
Angularity & Square form (for reflection/refraction) & 6\note{/4} & Product form (for reflection) & 4\note{/2} \\
 & $\eta_{i-1}^2 {{{\mbd}_{i}^2}} \left({({\mbd}_{i-1}\times {\mbn}_i) \cdot \mbb}\right)^2
-
\eta_i^2 {{{\mbd}_{i-1}^2}} \left({({\mbd}_{i}\times {\mbn}_i) \cdot \mbb}\right)^2=0
$ & &  $(\mbd_{i-1} \cdot {\mbn}_i) (\mbd_{i} \cdot {\mbt}_i) + (\mbd_{i-1} \cdot {\mbt}_i) (\mbd_{i} \cdot {\mbn}_i) = 0$  & \\
\bottomrule
\end{tabular}
\label{tab_equ}
\end{table*}

\subsection{Variable reduction}

Our key idea for variable reduction lies in expressing the barycentric coordinates of all specular vertices using that of the first vertex in a given specular chain. This is achieved recursively: \note{Each} vertex is represented by its preceding neighbor, as illustrated in Fig. \ref{fig_proj}(b). 

Specifically, we compute the intersection of the reflected/refracted ray with the $i\textrm{+}1$-th triangle $\mbT_{i+1}$ using the M{\"o}ller–Trumbore algorithm \cite{AkenineMller1997FastMS}, 
which represent $\mbu_{i+1}$ as a function of $\mbu_i$ and $\mbu_{i-1}$ in the following rational form named \textbf{rational coordinate mapping} in this paper:
\begin{equation}
\boxed{
\mbu_{i+1}(\mbu_i, \mbu_{i-1}) = \frac{{\left({\tilde{u}_{i+1}(\mbu_i, \mbu_{i-1})}, {\tilde{v}_{i+1}(\mbu_i, \mbu_{i-1})}\right)^\top}}{\kappa_{i+1}(\mbu_i, \mbu_{i-1})} ,
}
\label{eq_rct}
\end{equation}
where
\begin{equation}
\tilde{u}_{i+1}(\mbu_i, \mbu_{i-1})
=(
\tilde{\mbd}_i
\times
\mbe_{i+1, 2}
)\cdot
(\mbx_i - \mbp_{i+1, 0})
,
\end{equation}
\begin{equation}
\tilde{v}_{i+1}(\mbu_i, \mbu_{i-1})
=(
(\mbx_i - \mbp_{i+1, 0})
\times
\mbe_{i+1, 1}
)\cdot
\tilde{\mbd}_i,
\end{equation}
\begin{equation}
\kappa_{i+1}(\mbu_i, \mbu_{i-1})
=(
\tilde{\mbd}_i
\times
\mbe_{i+1, 2}
)\cdot
\mbe_{i+1, 1}.
\end{equation}
Here, $\tilde{\mbd}_i$ is a special and unnormalized version of $\hat{\mbd}_i$, which is determined by the scattering type at $\mbx_i$. To keep the expression of $\mbu_{i+1}$ rational, $\tilde{\mbd}_i$ is also expected to be rational. These rational expressions are beneficial for polynomialization since rational equations can be easily converted into polynomial equations.


\paragraph{Reflection}

The reflected direction at $\mbx_i$ can be written as
\begin{equation}
\hat{\mbd}_{i}=-2(\hat{\mbd}_{i-1}\cdot \hat{\mbn}_i)\hat{\mbn}_i+\hat{\mbd}_{i-1}.
\end{equation}
To eliminate the square roots in $\hat{\mbn}_i$ and $\hat{\mbd}_{i-1}$, \note{we introduce a scaled direction vector $\tilde{\mbd}_{i}$ through multiplying $\hat{\mbd}_{i}$ by ${\mbn}_i^2 \sqrt{\mbd_{i-1}^2}$:}
\begin{equation}
\tilde{\mbd}_{i} = 
-2(\mbd_{i-1} \cdot {\mbn}_i){\mbn}_i
+
\mbd_{i-1} {\mbn}_i^2,
\label{eq_431_di}
\end{equation}
which is a polynomial in $\mbu_{i-1}$ and $\mbu_i$.

\paragraph{Refraction}

The expression of the refracted direction at $\mbx_i$ is much more complicated:
\begin{equation}
\hat{\mbd}_i
=
{\eta_i'}
({\hat{\mbd}_{i-1} - (\hat{\mbd}_{i-1} \cdot \hat{\mbn}_i) \hat{\mbn}_i})
-
\sqrt{
1 - \eta_i'^2 (1-(\hat{\mbd}_{i-1} \cdot \hat{\mbn}_i)^2)
} \hat{\mbn}_i,
\end{equation}
where $\eta_i' = \eta_{i-1} / \eta_i$.
Similarily, we also multiply the above equation by $\mbn_i^2 \sqrt{\mbd_{i-1}^2}$, yielding
\begin{equation}
\tilde{\mbd}_i
=
\eta_i' (\mbd_{i-1} \mbn_i^2 - (\mbd_{i-1} \cdot \mbn_i) \mbn_i) - \sqrt{
\beta_i
}
\mbn_i,
\label{eq_refr_rational}
\end{equation}
where
\begin{equation}
\beta_i
=
\mbn_i^2 \mbd_{i-1}^2
- \eta_i'^2 \left(\mbn_i^2 \mbd_{i-1}^2
-
(\mbd_{i-1} \cdot \mbn_i)^2 \right).
\end{equation}
The square root operation in Eq. (\ref{eq_refr_rational}) forbids it to be rationalized accurately. 
Thus, we provide the following cheap piecewise rational approximation to $\sqrt{x}$ in the range of $[0,1]$. We subdivide the the range $[0,1]$ into 6 consecutive pieces. In each piece, $\sqrt{x}$ is fit by the following rational \note{function}:
\begin{equation}
 \frac{c_{0,i} + c_{1,i} x}{d_{0,i} + d_{1,i} x}, \, i=0,1,2...,5.
\end{equation}
The details of the coefficents $c_{0,i}, c_{1,i}, d_{i,0}, d_{1,i}$ are included in the supplemental material. The approximation provides an error less than $10^{-3}$, which bounds the angular error within a threshold \cite{Zeltner20, Wang20} and is sufficient for \note{most cases}.

\subsection{Bivariate specular polynomials}

\note{Until} now, for any specular chain, we can use rational coordinate mapping (Sec. 4.4) to express all $\mbu_i$ in the chain using $\mbu_1$. By putting them into the vertex constraint (Sec. 4.3) on the last vertex $\mbx_k$, we arrive at a bivariate form of specular polynomials:
\begin{equation}
\left\{
\begin{aligned}
&a(\mbu_1) = a(u_1, v_1) = 0, \\
&b(\mbu_1) = b(u_1, v_1) = 0,
\end{aligned}
\right.
\end{equation}
with $\mbu_1=(u_1, v_1)$ being the only variable\footnote{Again, $\mbu_0$ and $\mbu_{k+1}$ are dummy variables since $\mbx_0$ and $\mbx_{k+1}$ are fixed.}. This allows us to design practical solutions to simulate specular light transport with different types.
We show three special cases in what follows. The notations follow Heckbert's tradition \cite{Heckbert90}: $R$ denotes reflection and $T$ denotes refraction.


\paragraph{R} 
For specular reflection with a single bounce, \note{Eq. (\ref{eq_coplane}) and Eq. (\ref{eq_pf_result}) show that}
\begin{equation}
\left\{
\begin{aligned}
a(\mbu_1)=&
(
(\mbP_1 \mbu_1 - \mbx_0) \times
(\mbx_{2} - \mbx_{0})
) \cdot 
\mbN_1 \mbu_1
= 0, \\
b(\mbu_1) =& \left((\mbP_1 \mbu_1 - \mbx_0) \cdot \mbN_1 \mbu_1\right) \left((\mbx_2 - \mbP_1 \mbu_1) \cdot (\mbN_1 \mbu_1 \times \mbe_{1,1})\right)+  \\&\left((\mbx_2 - \mbP_1 \mbu_1) \cdot \mbN_1 \mbu_1\right) \left((\mbP_1 \mbu_1 - \mbx_0) \cdot (\mbN_1 \mbu_1 \times \mbe_{1,1})\right) = 0.
\end{aligned}
\right.
\end{equation}
The above two polynomials are of degree $2$ and $4$, respectively.

\paragraph{T}
The derivation of the refractive case is analogous\note{, using Eq. (\ref{eq_coplane}) and Eq. (\ref{eq_cpf_result}):}
\begin{equation}
\left\{
\begin{aligned}
a(\mbu_1)=&
(
(\mbP_1 \mbu_1 - \mbx_0) \times
(\mbx_{2} - \mbx_{0})
) \cdot 
\mbN_1 \mbu_1
= 0, \\
b(\mbu_1) =& \eta_0^2 (\mbx_2 - \mbP_1 \mbu_1)^2 \left( ((\mbP_1 \mbu_1 - \mbx_0) \times \mbN_{1} \mbu_1) \cdot \mbb \right)^2 - \\& \eta_1^2 (\mbP_1 \mbu_1 - \mbx_0)^2 \left( ((\mbx_2 - \mbP_1 \mbu_1) \times \mbN_{1} \mbu_1) \cdot \mbb \right)^2 = 0.
\end{aligned}
\right.
\end{equation}
The above two polynomials are of degree $2$ and $6$, respectively.

\paragraph{RR}
For the angularity relationship, we replace $\mbd_{i-1}$ in Eq. (\ref{eq_pf_result}) with $\tilde{\mbd}_{i-1}$ defined in Eq. (\ref{eq_431_di}). Combined with the coplanarity constraint, we obtain
\begin{equation}
\left\{
\begin{aligned}
&a(\mbu_1) = 
\left((\mbP_2 \mbu_2 - \mbP_1 \mbu_1) \times
(\mbx_{3} - \mbP_1 \mbu_{1}
)\right) \cdot 
\mbN_2 \mbu_2
= 0, \\
&b(\mbu_1) = 
(\tilde{\mbd}_{1} \cdot {\mbn}_2) (\mbd_{2} \cdot {\mbt}_2) + (\tilde{\mbd}_{1} \cdot {\mbt}_2) (\mbd_{2} \cdot {\mbn}_2) = 0.
\label{eq_rr1}
\end{aligned}
\right.
\end{equation}
Here, $\tilde{\mbd}_1$ is a polynomial in $\mbu_1$, while $\mbn_2$, $\mbd_2$, and $\mbt_2$ are all linear to $\mbu_2$. Recall that $\mbu_2$ is a rational expression of $\mbu_1$. Therefore, we can easily obtain polynomial systems in $\mbu_1$ from the above equations.

\paragraph{Other cases}
Table \ref{tab_sum} summarizes the above theoretical results. The first three lines list the multivariate formulations of specular polynomials, while the other lines show the bivariate versions for one or two reflection/refraction bounces via rational coordinate mapping. The generalization to bounces over two is straightforward. 


\begin{table}[tbp]
\centering

\caption{Summary of various forms of specular polynomials. Note that the time cost of solving a bivariate system is proportional to the cubic of the product degree of the two polynomials.}
\begin{tabular}{c|lccc}
\toprule
& \textbf{Type} & \textbf{Equation} & \textbf{\#Var.} & \textbf{Degree} \\ 
\midrule
& $R^k$ & Eqs. (\ref{eq_coplane}), (\ref{eq_pf_result}) & $2k$ & $2$, $4$ \\
\textbf{Multivar.} & $T^k$ & Eqs. (\ref{eq_coplane}), (\ref{eq_cpf_result}) & $2k$ & $2$, $6$ \\
& $(R|T)^k$ & Eq. (\ref{eq_ch}) & $3k+1$ & $2$ \\
\midrule
& $R$ & Eqs. (\ref{eq_coplane}), (\ref{eq_pf_result}) & 2 & 2, 4 \\
& $T$ & Eqs. (\ref{eq_coplane}), (\ref{eq_cpf_result}) & 2  & 2, 6 \\
\textbf{Bivar.} & $RR$ & Eqs. (\ref{eq_coplane}), (\ref{eq_pf_result}), (\ref{eq_rct}) & 2  & 10, 16 \\
& $RT,TR$ & Eqs. (\ref{eq_coplane}), (\ref{eq_cpf_result}), (\ref{eq_rct}) & 2  & 10, 24 \\
& $TT$ & Eqs. (\ref{eq_coplane}), (\ref{eq_pf_result}), (\ref{eq_rct}) & 2 & 18, 48 \\
\bottomrule
\end{tabular}
\label{tab_sum}
\end{table}

\section{An Efficient Polynomial Solver}

In this section, we develop a practical pipeline that effectively constructs and solves the aforementioned specular polynomials.

\subsection{Variable elimination using resultants}
Although we have converted specular constraints into simplified polynomials only containing two variables, finding the roots of the system is still challenging due to the high degree. 
However, many techniques exist in mathematical literature that can well solve the univariate polynomial with a high degree \cite{yukselHighPerformancePolynomialRoot2022, nakatsukasaVectorSpacesLinearizations2017}. This inspires us to further eliminate one variable in our bivariate specular polynomials.

In this paper, we resort to the hidden variable resultant method \cite{nakatsukasaComputingCommonZeros2015} due to its efficiency and robustness. Specifically, we eliminate one variable $u_1$, and convert the problem into finding the zeros of the B\'{e}zout resultant \cite{Bezout1779} $r(v_1)$, a univariate polynomial in terms of $v_1$. In this case, $v_1$ is called the hidden variable. 
The resultant $r(v_1)$ is defined as the determinant of the resultant matrix $\boldsymbol{R}(v_1)$ \cite{Stiller2004AnIT}:
\begin{equation}
r(v_1)= \det \boldsymbol{R}(v_1).
\end{equation} 
Here, for the B\'{e}zout resultant, the $(i,j)$-th element of the $n\times n$ matrix $\boldsymbol{R}(v_1)$ is given by\footnote{Different algorithms can be used for constructing the B\'{e}zout resultant matrix. In our implementation, we adopt a fast computation method \cite{chionhFastComputationBezout2002}. The pseudo-code is presented in the supplemental document.} \cite{Bezout1779}
\begin{equation}
\sum_{k=0}^{\min(i,n-1-j)} \left(a_{i-k}(v_1) b_{j+1+k}(v_1) - b_{i-k}(v_1) a_{j+1+k}(v_1)\right),
\label{eq_bezout}
\end{equation}
where $n$ is the degree of bivariate specular polynomials.
We call $r(v_1)$ the \textbf{univariate specular polynomial}. It connects to bivariate specular polynomials via $a_i(v_1)$ and $b_i(v_1)$, which represent the coefficient of $u_1^i$ in $a(u_1,v_1)$ and $b(u_1,v_1)$:
\begin{equation}
\left\{
\begin{aligned}
&a(u_1, v_1) = \sum_{i=0}^{n} a_i(v_1) u_1^i=0, \\
&b(u_1, v_1) = \sum_{i=0}^{n} b_i(v_1) u_1^i=0.
\end{aligned}
\right.
\end{equation}
This allows us to solve for one variable $v_1$ first and put it back into the original bivariate system to solve for the other variable $u_1$.
\note{We provide a running example to demonstrate how the resultant method works in the supplemental document.}

\subsection{Solving the univariate problem}

The last building block of our whole pipeline is an efficient solver to the univariate equation:
\begin{equation}\label{eq_univariate}
r(v_1) = \det \boldsymbol{R}(v_1) = 0.
\end{equation}
We perform Laplacian expansion for the determinant $\det \boldsymbol{R}(v_1)$ to obtain the coefficients of the resultant polynomial explicitly, and subsequently solve the univariate polynomial equations $r(v_1) = 0$.
Then we find the roots of univariate polynomials in a recursive manner: \note{The} derivative of a polynomial of degree $d$ is a polynomial of degree $d-1$, and the zeros of the latter \note{determine} the monotonic pieces of the former. On each monotonic piece, only a single root exists and can be found through a bisection solver. \note{Consistency is determined by checking whether the length of the bisection interval is smaller than a threshold, which is set to $10^{-9}$ in our implementation.}

However, a significant drawback of Laplacian expansion is its exponential time complexity, making it only practical in the case of a single bounce. For large matrices produced by two or more bounces, we found it not practical to obtain the resultant coefficients explicitly. Instead, we opt to directly find the zeros by uniformly dividing $[0,1]$ and running bisection on each piece of which the determinant on the two endpoints are with different signs. We evaluate the determinant using Gaussian elimination and find that $100$ pieces and $10$ bisection iterations work well in our test scenes.



\paragraph{Discussion}
The piecewise bisection solver requires a large number of pieces in the case that solutions are clustered. However, our experiments show that the errors are negligible and have little influence on our test scenes.
An alternative and theoretically comprehensive approach is by employing linearization \cite{Golub2012MatrixComputation}, which converts the above root-finding problem into a generalized eigenvalue problem. In mathematics, this can be well solved by QZ decomposition.
Unfortunately, the computational burden of QZ decomposition is prohibitively high since the time complexity is $\mathcal{O}(kn^6)$, where $n$ is the degree of the bivariate polynomial system and $k$ is the number of iterations. Developing a fast and accurate solver to Eq. (\ref{eq_univariate}) with an arbitrarily high degree is still an open problem in mathematics.



 


\subsection{The whole pipeline}

\begin{figure}
    \centering
    \includegraphics[width=\linewidth]{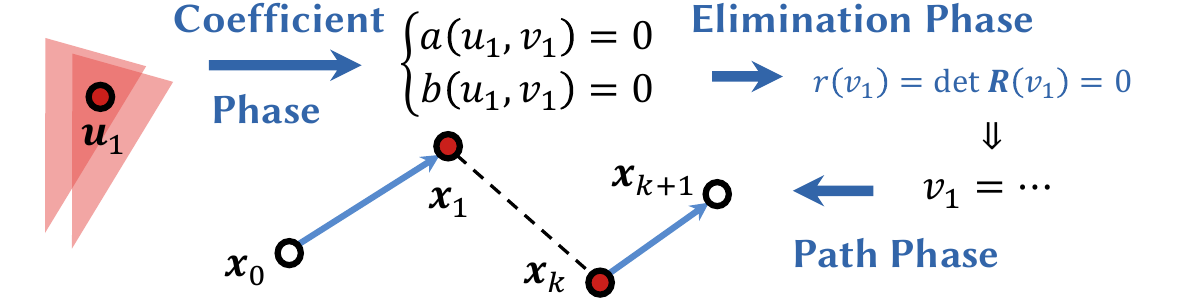}

    \caption{Overview of the pipeline. Taking vertex positions and normals as inputs, our pipeline systematically constructs specular polynomials, converts the multivariate systems into a univariate problem via constructing resultant matrices, and subsequently solves the univariate problem. Finally, it validates the solutions and generates admissible paths.}
    \label{fig_pipe}
\end{figure}

In summary, the whole pipeline of our method consists of three phases shown in Fig. \ref{fig_pipe}. \note{Please refer to the supplemental document for more details and pseudo-code snippets.}

\paragraph{Coefficient phase}

We first convert the vertex positions and normals of triangles along a specular chain into bivariate polynomials on $u_1$ and $v_1$, according to the steps described in Sec. 4. The major task in this phase is to determine the coefficients of each specular polynomial.

\paragraph{Elimination phase}

This phase converts the bivariate system on $u_1$ and $v_1$ into a univariate problem using the hidden variable resultant method introduced in Sec. 5.1. Here, each element of the resultant matrix $\boldsymbol{R}(v_1)$ is now a univariate polynomial in $v_1$. We only need to find all zeros of the determinant of the resultant matrix $\det \boldsymbol{R}(v_1)$. Then, we find all zeros of the matrix polynomial determinant using the method introduced in Sec. 5.2.


\paragraph{Path phase}

As we have determined the solutions for $v_1$, we reintegrate them into the original polynomial $a(u_1, v_1)$ and solve for the other variable $u_1$, which only requires solving an univariate equation $a(u_1)|_{v_1}$ with explicitly known coefficients. Finally, we generate the path utilizing the solved barycentric coordinates, check the specular constraints in the path space \cite{Wang20, Zeltner20}, and conduct the visibility test. An admissible path will be rejected when the ray from any vertex $\mbx_i$ towards the next vertex $\mbx_{i+1}$ is blocked. If all the checks pass, our pipeline reports a valid solution and evaluates its contribution \cite{Wang20, Jakob12}. \note{Note that when there are multiple connections between two separators, we sum the contributions from all of them.}

\paragraph{Complexity analysis}

\note{
The time complexity for solving all chains given two separators is $\mathcal{O}(t(n^4+Cn^3))$. Here, $C$ is the product of the number of intervals and bisection iterations. $t$ is the number of triangle tuples, which depends on pruning techniques \cite{Walter09, Wang20}. $n$ is the degree of bivariate polynomials, which is approximately $4^k 2^r$ with $r$ being the number of refractive vertices in the specular chain. The space complexity is $\mathcal{O}(n^3+C)$. Please refer to the supplemental material for a detailed discussion.
}

\begin{figure}
\centering
\includegraphics[width=\linewidth]{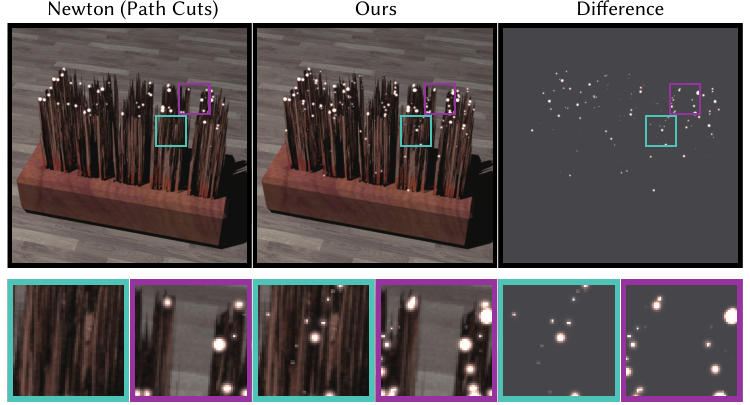}  
    \includegraphics[width=\linewidth]{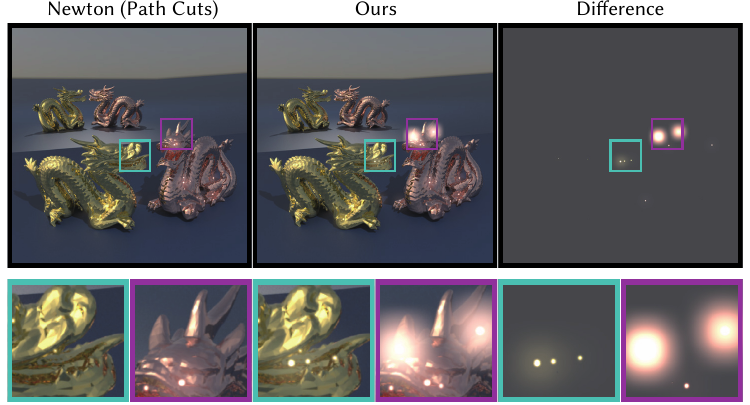}
\includegraphics[width=\linewidth]{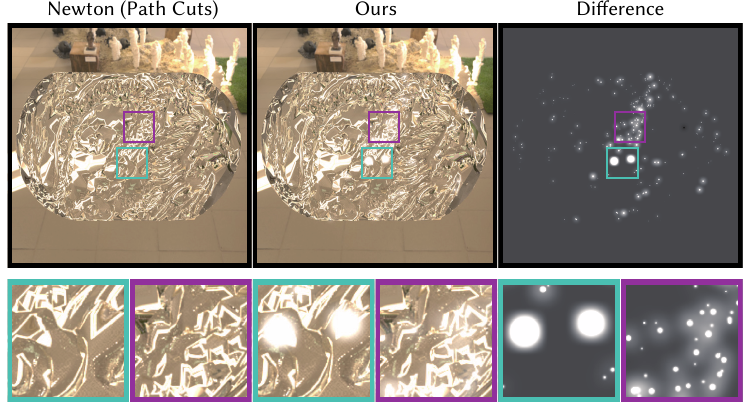}
    \caption{Glints rendering featuring different types of specular chains. \textbf{Top}: \textit{R} chains on a metal brush. \textbf{Middle}: \textit{RR} chains in a scene with two specular dragons and a mirror. \textbf{Bottom}: \textit{TT} chains passing through a \note{relief, for which we also show glint-only images in the supplemental material.}}
    \label{fig_glints}
\end{figure}

\section{Results}

We have implemented our method on top of the Mitsuba renderer \cite{Mitsuba}, and apply it on both glints rendering and caustics rendering. In the current implementation, we run our solver to find admissible specular paths for each tuple of triangles \note{that passed the pruning of non-contributing tuples following \citet{Wang20}, which is based on a path space hierarchy and interval arithmetic bounds.} For caustics rendering, we integrate our method into a conventional path tracer and perform our solver independently for each pair of separators. We sample the separators similarly to \note{\citet{Fan23MPG, Zeltner20}}. 
The reference images of caustics rendering are generated by \note{the unbiased variant of} Specular Manifold Sampling (SMS) \cite{Zeltner20} with a very high sample rate. 
All timing measurements were conducted on a PC with a Core i9-13900KF processor and an RTX 4080 graphics card.

\subsection{Glints rendering}

In this task, we aim to find all admissible specular chains connecting a camera sample and a light sample in a deterministic way. We take Path Cuts \cite{Wang20}, another deterministic method based on Newton iterations, as the competitor. The rendered images are postprocessed by a bloom filter, to better show the high-frequency glints.
As a state-of-the-art method in simulating pure specular paths, Path Cuts can find many challenging paths that are not affordable by conventional path tracers. However, it still runs the risk of missing some important admissible paths, since Newton's method involved in Path Cuts \note{uses the center of triangles to construct seed paths and} does not always converge, resulting in loss of glints as highlighted in the difference maps of Fig. \ref{fig_glints}. In these figures, we show glints stemming from different types of specular chains.

\begin{figure*}
    \centering
    \includegraphics[width=1\textwidth]{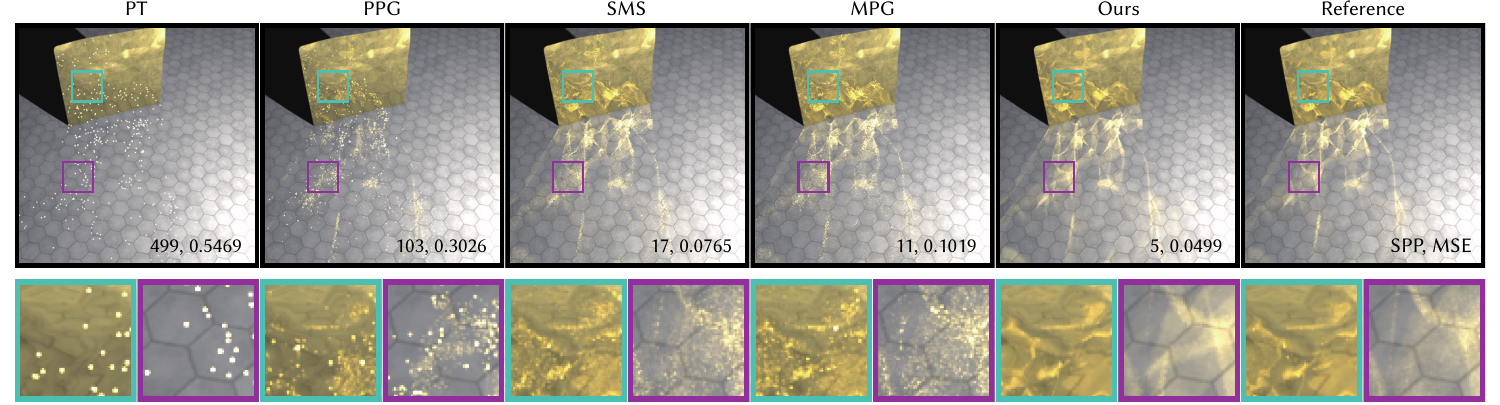}
    \includegraphics[width=1\textwidth]{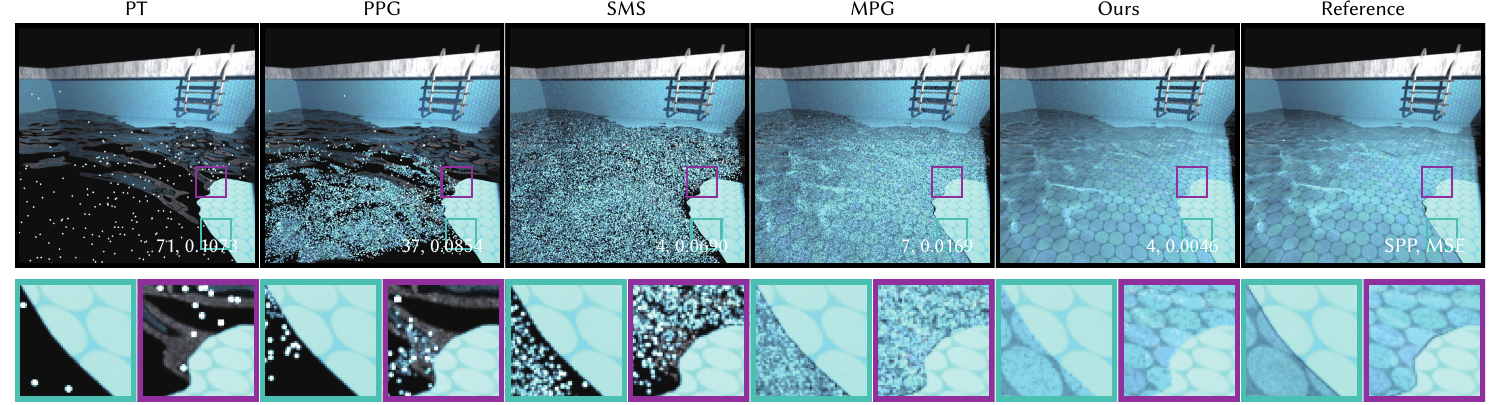}
    
    \caption{Equal\note{-}time comparisons (10 sec) on caustics rendering \note{with Path Tracing (PT) \cite{Kajiya86}, Practical Path Guiding (PPG) \cite{Muller17}, \note{the unbiased variant of} Specular Manifold Sampling (SMS) \cite{Zeltner20} and Manifold Path Guiding (MPG) \cite{Fan23MPG}} on the Plane and Pool scenes. We report the samples per pixel \note{(SPP)} and quantitative errors in terms of mean square error \note{(MSE)}. }
    \label{fig_caustics_main}
\end{figure*}

\begin{figure}
    \centering
    \includegraphics[width=1\linewidth]{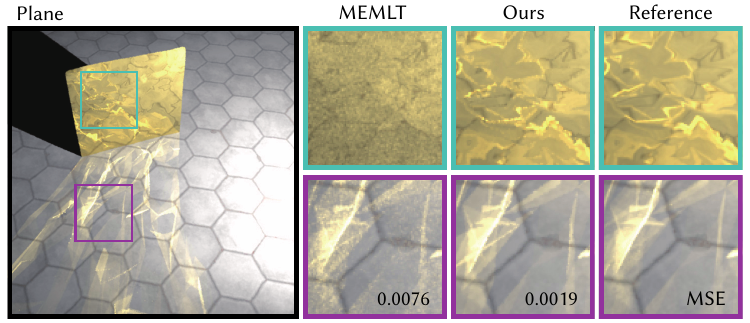}
    \includegraphics[width=1\linewidth]{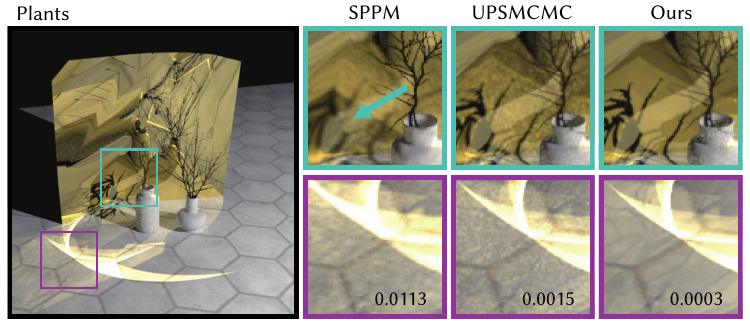}
    \caption{\textbf{Top}: Equal\note{-}time (30 sec) comparison with \note{Manifold Exploration Metropolis Light Transport (MEMLT) \cite{Jakob12}}. \textbf{Bottom}: Equal\note{-}time comparison (10 sec) with \note{Stochastic Progressive Photon Mapping (SPPM) \cite{Hachisuka09SPPM} and Metropolised Bidirectional Estimator (UPSMCMC) \cite{Sik2016RobustLT}}.}
    \label{fig_photon}
\end{figure}

Instead, our method utilizes comprehensive information about each tuple of triangles along a specular path, converting the challenging light transport problem losslessly into a polynomial root-finding problem. Using the numerical tools discussed above, our method succeeds in producing rendering results that include significantly more glints than the existing method in our test scenes. 

We do not compare our method to the interval Newton's method \cite{Mitchell92, Walter09} since it requires several orders of magnitudes more time than the \note{non-interval} one, making it impractical for most scenes \cite{Wang20}. Instead, through polynomialization and variable reduction, we can solve the original problem very efficiently. The run-time cost of our method is generally in the same order of magnitude as the regular Newton's method, and is even faster than Newton's method in the case of a single bounce, while simultaneously finding more solutions.

\subsection{Caustics rendering}

For caustic rendering, we compare our method to conventional Path Tracing (PT) \cite{Kajiya86}, Practical Path Guiding (PPG) \cite{Muller17}, \note{the unbiased variant of} Specular Manifold Sampling (SMS) \cite{Zeltner20} and Manifold Path Guiding (MPG) \cite{Fan23MPG}. In particular, SMS and MPG are the state-of-the-art methods for caustics rendering. Unlike ours, both methods involve stochastic sampling to search specular chains. We show equal-time comparisons in Fig. \ref{fig_caustics_main}.

Traditional guiding approaches face challenges when confronted with specular interactions due to the high-frequency variations inherent in the radiance distribution. Visual comparisons distinctly reveal the limitations of a typical guided path tracer, such as PPG, in handling complex light paths featuring tiny area light and specular vertices, resulting in outliers and energy loss. Similar issues will also happen with more advanced guiding approaches \cite{Rath2020VarianceawarePG, Ruppert2020RobustFO, dodikPathGuidingUsing2022, Rath23}. 

SMS often exhibits noticeable noise, primarily due to its uniform sampling of seed chains.  
Moreover, an important issue with manifold-based methods is that they have an unbounded probability of finding a solution, which may be very small and lead to significantly high variance. This becomes particularly apparent in regions associated with admissible chains featuring extremely small convergence basins, such as the boundary of the water in the pool. 

While MPG mitigates this issue by employing importance sampling of seed chains, the learned distribution requires a fairly long time to become accurate enough to perfectly fit the target function. As a result, some regions in the image rendered using MPG still exhibit high variance and outliers. Additionally, for manifold-based methods, the introduction of reciprocal probability estimation contributes to variance. Consequently, even in simple scenes (e.g., Plane) where the importance distributions are easy to fit, the rendering result is still with visible noise.

Our method is free from stochastic sampling, thus avoiding the noise and outliers faced by manifold-based methods. Consequently, our method generates results with very low variance and succeeds in finding all solutions in both two scenes.

In Fig. \ref{fig_photon}, we compare our method with Manifold Exploration Metropolis Light Transport (MEMLT) \cite{Jakob12, Veach97MLT}. Admissible chains are required by MEMLT as their seed paths. For SDS paths, it is inefficient to just rely on PT or BDPT for finding seed paths. This leads to either over-bright artifacts when Markov chains become trapped in small regions, or energy loss if no seed path exists for a given region. Working as a deterministic method integrated into a standard Monte Carlo sampling framework such as path tracing, our approach eliminates blotchy energy loss in the images.

\begin{figure}
    \centering
    \includegraphics[width=\linewidth]{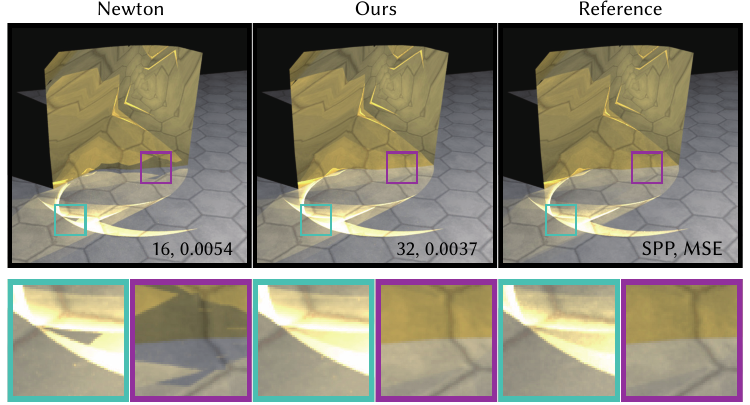}
    \includegraphics[width=\linewidth]{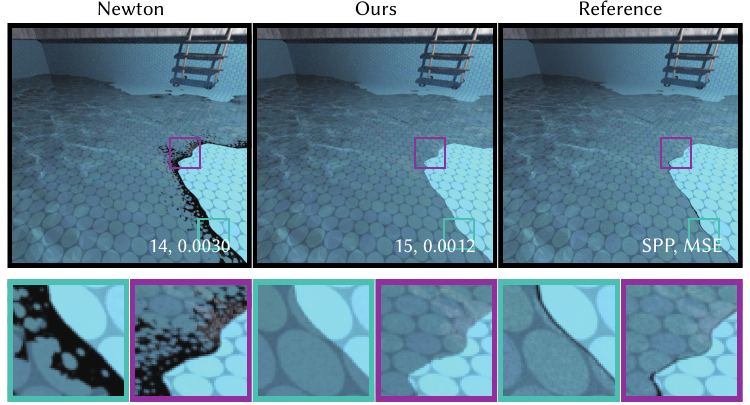}
    \caption{Equal\note{-}time (30 sec) comparison between Newton's method and our specular polynomials on caustics rendering.}
    \label{fig_caustics_pathcuts}
\end{figure}

Additionally, we compare our method with photon-based approaches, in particular, Stochastic Progressive Photon Mapping (SPPM) \cite{Hachisuka09SPPM} and Metropolised Bidirectional Estimator (UPSMCMC) \cite{Sik2016RobustLT}. 
For SPPM, we reduce the photon lookup radius to achieve a balance between bias and variance.
Besides visible noise, SPPM will cause over-blurriness (e.g., the cyan inset) due to the nature of density estimation. This problem also exists in UPSMCMC, which struggles to preserve the sharp edge of shadows in the cyan inset and simultaneously produces noticeable noise. 
In contrast, our deterministic approach solves for admissible paths directly and accurately, thus avoiding the issues of density estimation and yields almost noise-free rendering with detailed patterns of caustics.


\subsection{Validations}

\paragraph{Newton's method \note{vs.} specular polynomials on caustics rendering}

To further verify that our deterministic method works well on caustics rendering, we replace our solver based on specular polynomials with the regular Newton's method. We adopt the Path Cuts framework to ensure that no stochastic sampling is involved, and the seed chain of Newton's method is heuristically generated by connecting the center of each triangle \cite{Wang20}. Therefore, both methods work in a deterministic way.
From the comparison in Fig. \ref{fig_caustics_pathcuts}, we see that regular Newton's method with deterministic heuristic initialization is inadequate in identifying all solutions when a tuple of triangles encompasses multiple admissible paths. If the seed significantly deviates from the solution, Newton's method will diverge. Consequently, the rendering results suffer from energy loss and visual artifacts (e.g., the bottom of the plane), as shown in Fig. \ref{fig_caustics_pathcuts}. In comparison, our method finds all the admissible paths in our test scenes, producing results that are nearly identical to the reference images.

\begin{figure}[tbp]
    \centering
    \includegraphics[width=1\linewidth]{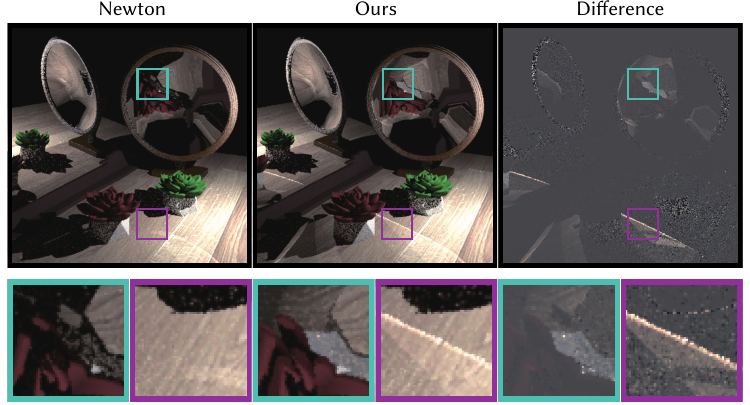}
    \caption{Equal\note{-}sample (1 spp) comparison of caustics rendering featuring specular chains of \textit{RR} type.}
    \label{fig_rr_cau}
\end{figure}

We also apply our method to render caustics featuring specular chains of length two in Fig. \ref{fig_rr_cau}. Since the dimensionality is high, the convergence basin of each solution for Newton's method may become extremely small. Simply relying on a heuristic initialization hardly finds the admissible paths in some regions, missing many caustics. Again, our method, using the bisection-based solver, works well, yielding caustics with complete shape and fine details.

\begin{figure}
    \centering
    \includegraphics[width=1\linewidth]{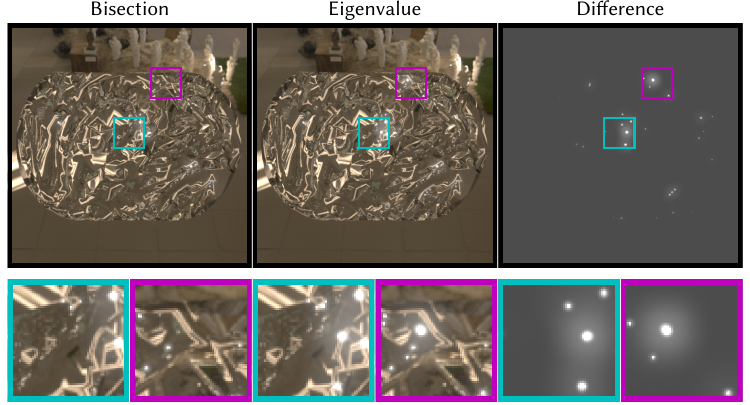}
    \caption{Comparison of the bisection and eigenvalue solver. The eigenvalue solver uses \note{275$\times$} more time than our bisection solver.}
    \label{fig_val_eig}
\end{figure}

\paragraph{Bisection solver \note{vs.} eigenvalue solver}

In Fig. \ref{fig_val_eig}, we compare our direct solver using bisection with an eigenvalue-based solver using QZ decompositions. Eigenvalue solver in theory can ensure global convergence of the polynomial eigenvalue problem. However, its computational burden is extremely unaffordable. Our bisection-based solver finds almost all the solutions, generating rendering results of the same quality as QZ, consuming significantly less time.

\paragraph{Impact of the number of pieces for bisection}

\begin{figure}
    \centering
    \includegraphics[width=1\linewidth]{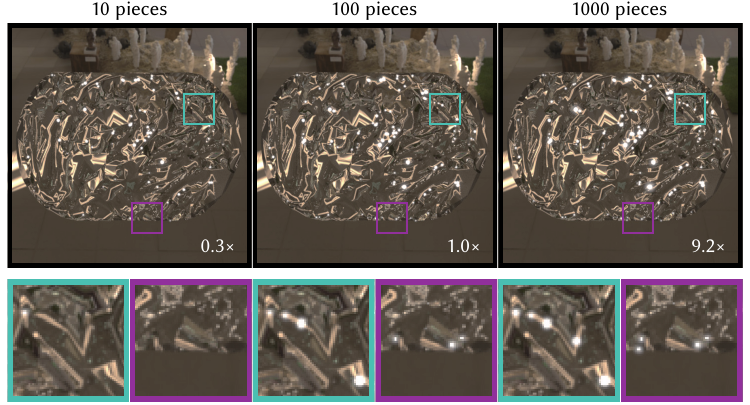}
    \caption{Impact of the number of pieces in the bisection solver. \note{The number shows the relative running time of our solver. Our choice of 100 pieces achieves an appropriate balance between accuracy and speed.}}
    \label{fig_val_bis}
\end{figure}

Additionally, we verify the influence of the pieces on the bisection solver in Fig. \ref{fig_val_bis}. \note{
Using 10 pieces misses many solutions, whereas 1000 pieces greatly reduces performance.} Our choice of $100$ pieces works well in this complex scene of glints rendering.

\subsection{Performance}

In Table \ref{tab_perf}, we report the rendering statistics of glints rendering. 
For specular chains with type $R$, our CPU-based solver is $3.3\times$ faster than regular Newton iterations. For type $T$, ours is also $2.5\times$ faster. It is important to note that in both cases, the bottleneck, which takes nearly half of the time, is spent on finding the root of the univariate polynomial, which is dependent on the choice of the underlying solver. The construction of our bivariate specular polynomials and the conversion to univariate polynomials only takes a quite small amount of time.

For two specular bounces, we develop \note{an} efficient GPU implementation. Most parts of our method are polynomial arithmetics and determinant evaluations, which is GPU-friendly since the computation steps of each thread are the same. Thanks to the strong parallelization capability of modern GPU, our method is slower than the regular Newton's method by only one order of magnitude, while finding significantly more solutions. 

\section{Discussions and Limitations}

\paragraph{Accurate rational coordinate mapping for refraction}

\begin{figure}
    \centering  \includegraphics[width=\linewidth]{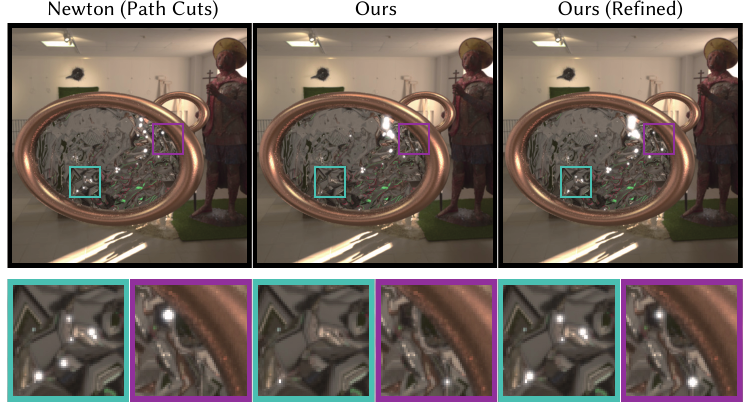}
    \caption{\note{Limitations of the rational approximation for refraction. Small differences in the refracted direction amplify when the specular surfaces are far away from each other. Yet, simply refining our solutions with only one iteration of Newton's method leads to much better accuracy.}}
    \label{fig_lim}
\end{figure}

When dealing with refraction, polynomials naturally have limitations. It is challenging to polynomialize the square roots in the refracted direction expression. To address this problem, we employ first-order rational approximation. Nonetheless, \note{this may lead to inaccuracy as shown in Fig. \ref{fig_lim}}. Future works are required to find better approaches.

\begin{table}[t]
    \caption{Time usage (${\upmu}$s) per triangle of Newton's method and our solver measured on glints rendering scenes. We show the time usage of constructing bivariate polynomials and the matrix $\boldsymbol{R}$ (Poly.), expanding determinants (Det.), and solving for $v_1$ and $u_1$ (Sol.). \note{The timing for the bisection solver is divided into choosing intervals (Det.) and performing bisections (Sol. $v_1$).}}
    \centering
    \begin{tabular}{l|cccc|c|c}
    \toprule
       Type & \multicolumn{5}{c|}{Ours}          & Newton \\
            & Poly. & Det. & Sol. \note{$v_1$} & Sol. \note{$u_1$} & Total & Total \\
       \midrule
\textit{R} \note{(CPU)} & 0.039 & 0.155 & 0.234 & 0.121 & 0.549 & 1.817 \\
\textit{T} \note{(CPU)}& 0.221 & 0.329 & 0.283 & 0.157 & 0.990 & 2.427\\
\textit{RR} \note{(GPU)}& \note{0.649} & \note{1.617} & 1.931 & 0.739 & 4.936 & \note{0.306} \\ 
 \bottomrule
    \end{tabular}
    \label{tab_perf}
\end{table}

\paragraph{Better numerical methods for root-finding}

\begin{figure}
    \centering  \includegraphics[width=\linewidth]{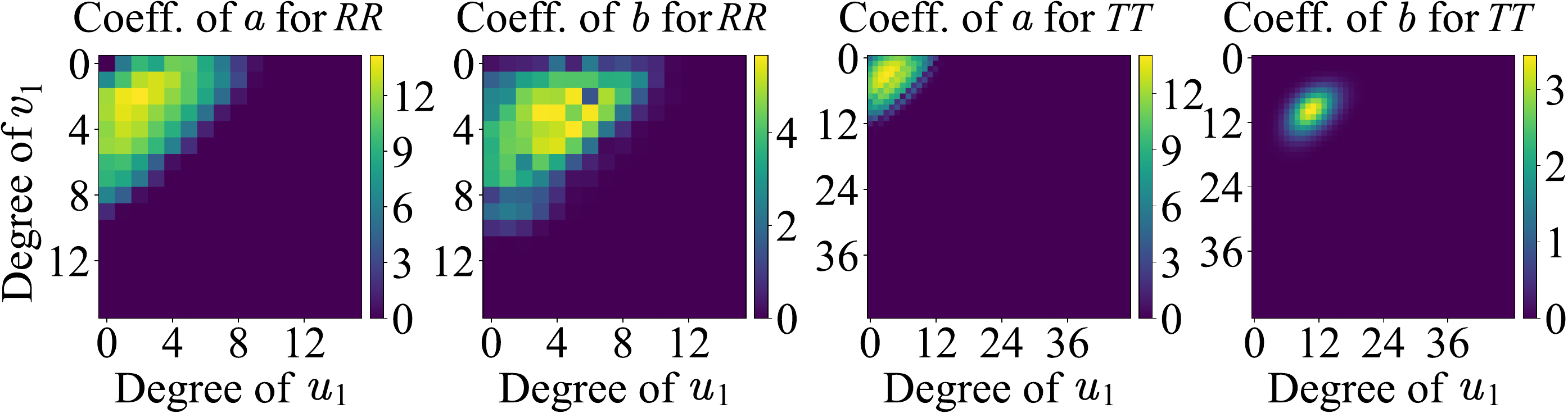}
    \caption{Examples of the coefficients of bivariate specular polynomials. We show the absolute value of coefficients.}
    \label{fig_bivar}
\end{figure}

Our primary focus in this work revolves around the derivation and simplification of specular polynomials. The bisection solver we used may find fewer solutions than the slow but comprehensive eigenvalue solver.
Nevertheless, finding accurate and comprehensive solutions to high-order polynomial equations poses challenges due to their high time complexity and numerical instability. Notably, our polynomial systems exhibit specific characteristics. For instance, the coefficients are predominantly large only for low-order terms, as shown in Fig. \ref{fig_bivar}. Employing better numerical tools tailored for such tasks remains a prospect for future exploration.

\paragraph{Long specular chains}

While we primarily investigate small specular chains, our approach can also be applied to longer ones. However, because of the combinatorial explosion issue, it is not feasible to find all the specular chains in these situations. Combining our method with some stochastic method would be useful since unbiased specular transport involving three or more specular vertices can currently only be resolved through stochastic sampling utilizing reconstructed importance distributions.

\paragraph{Theoretical unbiasedness}

Our specular polynomials are accurate \note{except for the approximated rational expansion of refraction.} However, comprehensive solving for two or more bounces requires a computationally expensive eigenvalue solver or a bisection solver with the number of pieces that tend to infinity. In our test scenes, our finite piece bisection solution performs well. Nonetheless, integrating a stochastic approach can be helpful if one wishes to guarantee \note{unbiasedness} \cite{Zeltner20,misso22}.

\paragraph{Superfluous solutions}

\note{The use of the resultant method introduces superfluous solutions, i.e., solutions of the univariate polynomials may not correspond to a valid specular path. For example, in the brush scene of Fig. \ref{fig_glints}, 52285 solutions of resultants are found, but only 6710 (13\%) passes the check of path space constraints. However, this only influences the performance of solving for $v_1$, which is a relatively small part of the total computation time.}

\paragraph{Handling near-specular vertices}

Extending our method to support glossy chains is straightforward. After sampling the normal offset for glossy vertices, the admissible chains corresponding to the offset remain finite, and the problem reduces to pure specular situations. Thus, we sample the offset normal from the microfacet distribution before solving the specular chains connecting two separators, as in previous work \cite{Kaplanyan2014, Hanika15, Zeltner20, Fan23MPG}. 

\paragraph{Surface representations}

\note{We have derived specular polynomials for triangles with interpolated normals, which are commonly utilized in existing rendering pipelines. Supporting other surface representations is a good avenue for future works.}

\section{Conclusion}

Specular light transport, with the underlying multivariate root-finding problem, is a long-standing challenge for physically-based rendering. Existing methods simply perform Newton's method in the high-dimensional path space, which heavily relies on a proper seed sampling strategy and easily introduces high variance or bias. 

As a fundamentally new methodology, specular polynomials proposed in this paper reformulate this problem into a univariate polynomial root-finding problem on a given interval, by applying polynomialization of vertex constraints, rational coordinate mapping, and the hidden variable resultant method. This makes the challenging light transport problem tractable and offers many benefits when integrated into existing rendering pipelines.
Evaluations on the rendering of glints and caustics demonstrate our superiority in terms of accuracy and performance when compared to prior Newton iteration-based solutions. 

We believe our work has made an important advance in dealing with unbounded convergence of Monte Carlo rendering and stochastic sampling for specular light transport, which may open up new research avenues for rendering intricate optical effects.

\begin{acks}
We would like to thank the anonymous reviewers for their valuable suggestions. 
This work was supported by the National Natural Science Foundation of China (No. 61972194 and No. 62032011).

\end{acks}

\bibliographystyle{ACM-Reference-Format}
\bibliography{mainref}

\end{document}